\documentstyle[emulateapj,epsf]{article}
\submitted{To appear in AJ, January 2001}
\lefthead{Blakeslee, Metzger, Kuntschner, \& C\^ot\'e}
\righthead{Lensing in Hercules}
\def\lta{\mathrel{\rlap{\lower 3pt\hbox{$\mathchar"218$}}
     \raise 2.0pt\hbox{$\mathchar"13C$}}}
\def\gta{\mathrel{\rlap{\lower 3pt\hbox{$\mathchar"218$}}
     \raise 2.0pt\hbox{$\mathchar"13E$}}}
\def\kms{km~s$^{-1}$}

\def\hMpc{$h^{-1}\,$Mpc}
\def\hkpc{$h^{-1}\,$kpc}
\def\etal{{et~al.}} 
\def\pf{\ifmmode{{\hbox{\sc psf}}}\else{{\sc psf}}\fi}
\def\mM{\ifmmode(m{-}M)\else$(m{-}M)$\fi}
\def\msun{\ifmmode{\hbox{M$_\odot$}}\else{M$_\odot$}\fi}
\def\scl{\ifmmode{\sigma_{\rm cl}}\else{$\sigma_{\rm cl}$}\fi}

\begin{document}
\tighten

\title{Lensing in the Hercules Supercluster\altaffilmark{1}}

\author{John P. Blakeslee\altaffilmark{2,3,4}, Mark R. Metzger\altaffilmark{3},
Harald Kuntschner\altaffilmark{2}, and Patrick C\^ot\'e\altaffilmark{3,5,6}}
\authoremail{J.P.Blakeslee@durham.ac.uk}
\authoremail{mrm@grus.caltech.edu}
\authoremail{Harald.Kuntschner@durham.ac.uk}
\authoremail{pc@astro.caltech.edu}
 
\altaffiltext{1}{Based on observations obtained at the W.M. Keck Observatory,
which is jointly operated by the California Institute of Technology, the
University of California, and the National Aeronautics and Space Administration.}
\altaffiltext{2}{Department of Physics, University of Durham, South Road, 
Durham, DH1 3LE, United Kingdom}
\altaffiltext{3}{California Institute of Technology,
Mail Stop 105-24, Pasadena,~CA~91125}
\altaffiltext{4}{Current address: Department of Physics and Astronomy,
Johns Hopkins University, Baltimore, MD 21218}
\altaffiltext{5}{Sherman M.\ Fairchild Fellow}
\altaffiltext{6}{Current address: Department of Physics and Astronomy,
Rutger University, 136 Frelinghuysen Road, Piscataway NJ, 08854-8019}

\begin{abstract}
We report Keck LRIS observations of an arc-like background galaxy near the
center of Abell~2152 ($z=0.043$), one of the three clusters 
comprising the Hercules supercluster.  
The background object has a redshift $z=0.1423$ and is situated 25\arcsec\
north of the primary component of the A2152 brightest cluster galaxy (BCG).
The object is about 15\arcsec\ in total length and has a reddening-corrected
$R$-band magnitude of $m_R = 18.55\pm0.03$.  Its spectrum shows numerous
strong emission lines, as well as absorption features.
The strength of the H$\alpha$ emission would imply a star formation rate 
$\hbox{SFR}\approx3\;h^{-2}\,\msun\,$yr$^{-1}$ in the absence of any lensing.
However, the curved shaped of this object and its tangential orientation
along the major axis of the BCG suggest lensing.  We model the A2152 core
mass distribution including the two BCG components and the cluster potential.
We present velocity and velocity dispersion profile measurements for the
two BCG components and use these to help constrain the potential.  
The lens modeling indicates a likely magnification factor of $\sim1.9$ 
for the lensed galaxy, making A2152 the nearest cluster in which such significant
lensing of a background source has been observed.
Finally, we see evidence for a concentration of early-type galaxies at
$z=0.13$ near the centroid of the X-ray emission previously attributed to
A2152.  We suggest that emission from this background concentration is the
cause of the offset of the X-ray center from the A2152 BCG.  The background
concentration and the dispersed mass of the Hercules supercluster could add
further to the lensing strength of the A2152 cluster.  
\end{abstract}


\section{Introduction}

Strong gravitational lensing of background sources by clusters of galaxies
provides some of the most unambiguous evidence for the presence of large
amounts of dark matter in clusters.  Lensing studies suggest that the mass
in clusters is greater and more centrally concentrated than implied, for
instance, by the X-ray properties (e.g., Mellier, Fort, \& Kneib 1993;
LeF\`evre \etal\ 1994; Waxman \& Miralda-Escude 1995; Smail \etal\ 1995).
Moreover, the mass appears to clump around the luminous galaxies, so that
consideration of substructure and the effect of the massive central galaxy
are often important for understanding the observations (e.g.,
Miralda-Escud\'e 1995; see also Tyson, Kochanski, \& Dell Antonio 1998).
This is particularly true in more moderate mass clusters with velocity
dispersions $\scl \lta 1000$ \kms, for which the proportionally larger
gravitational effects of a massive cD must be included to properly account
for the lensing (Williams, Navarro, \& Bartelmann 1999).

Although there have been great advances in our knowledge of the form of
the density distributions of massive dissipationless particles in 
simulations of various resolution (e.g., Dubinski \& Carlberg 1991; Navarro, 
Frenk, \& White 1996, 1997; Moore \etal\ 1998, 1999b), our knowledge of the
central mass distributions of actual galaxy clusters remains fragmentary.
Low redshift clusters, because of their larger angular sizes, can in
principle provide a better opportunity for studying these mass distributions.
Their lensing properties are acutely sensitive to the shape of the inner
cluster potential, although because such clusters tend to be of modest mass
% colossal, gigantic, huge, enormous, massive, prodigious, high mass
(very high-mass clusters being few and far between), the individual galaxies
will produce relatively larger perturbations on this potential, complicating
the interpretation. 

Until recently, moderate- to high-redshift clusters have been the
exclusive purview of strong lensing studies, but lensing by
low-redshift clusters has been gaining increased attention.
For instance, Allen, Fabian, \& Kneib (1996)
discovered a redshift $z=0.43$ lensed arc in the massive cooling 
flow cluster PKS\,0745--191 at $z=0.103$.
Campusano, Kneib, \& Hardy (1998) explored lensing models for a
$z=0.073$ spiral galaxy located near the central elliptical in
the richness class~0 cluster A3408 at $z=0.042$; they obtained an
upper limit to the magnification factor of $\sim$1.7 from their
``maximum mass'' model.  This object turned out to be the only
viable lensing candidate found in an imaging survey of 33 southern
galaxy clusters with $z\le0.076$ by Cypriano \etal\ (2000), who
also presented an estimate of the expected number of arcs and
arclets in low-redshift clusters.  
Blakeslee \& Metzger (1999) discovered a lensed arc at $z=0.573$ 
in the nearby cD cluster A2124 at $z=0.066$ and found 
magnification factors near 10 from their best-fitting models 
of this system.

Despite the recent progress, the lensing properties of low-redshift clusters
remain unclear.  For instance, A2124 had by far the lowest redshift in the
sample studied by Williams \etal\ (1999) of 24 lensing clusters with
measured velocity dispersions, the next nearest being at $z=0.171$.  In
this paper, we present imaging and spectroscopic observations and
give a lensing analysis of the redshift $z{\,=\,}0.043$ cluster Abell 2152
in the Hercules supercluster.

\section{Cluster Properties}
\label{sec:clusprops}

The Hercules supercluster is a close grouping of three Abell clusters:
the richness class~2 cluster A2151 (the classical ``Hercules cluster'')
and the two richness~1 clusters A2147 and A2152, all at $z\sim0.04$.
The supercluster was first identified by Shapley (1934). 
Although A2151 dominates from the standpoint of the number of galaxies, 
A2152 is projected closest to the center of the grouping, which
is so tight that a circle of just $2^\circ$ ($\,\lta\,$4\,\hMpc)
in diameter easily encompasses all three clusters. 
Two other Abell clusters, A2148 and A2107, lie at
about the same redshift and are only $\sim 15$~\hMpc\ away.  
At lower density enhancements, the Hercules supercluster forms part of 
an extended, filamentary supercluster that includes A2162 and A2197/A2199
about 3000~\kms\ in the foreground and A2052/A2063 about 3000~\kms\ in
the background (Abell 1961; Postman, Huchra, \& Geller 1992). 
This extended ten-cluster supercluster includes more members than any
other supercluster in the Postman \etal\ (1992) catalogue.

All three members of the Hercules supercluster proper 
% (A2147, A2151, and A2152)
are classified as Bautz-Morgan type~III, Rood-Sastry class F 
clusters (Abell \etal\ 1989; Struble \& Rood 1987), indicating 
morphological irregularity and perhaps dynamical youth.  All three
have relatively high spiral fractions, with that of A2151 being near 50\%
(Tarenghi \etal\ 1980). A2151 is also famous for its
high degree of internal subclustering (e.g., Bird \etal\ 1995).
A2152, the focus of this paper, has two bright early-type galaxies
at its center (see Figure~\ref{fig:img}) that together are designated
UGC\,10187.  The brighter component (NED~02) was chosen
by Postman \& Lauer (1995) as the A2152 brightest cluster galaxy (BCG) and
has a redshift $z=0.0441$.  The other component (NED~01) is 47\arcsec\
to the northwest and about 0.5\,mag fainter; Postman \& Lauer classify it
as the second ranked galaxy (SRG). The SRG has a relative velocity
of $+$330 \kms\ with respect to the BCG (see \S\ref{sec:kinematics}).

Because A2152 and A2147 are separated by just 1.8~\hMpc, measurements of
their velocity dispersions are problematic.  Zabludoff \etal\ (1993) reported
$\scl = 1081$ \kms\ for A2147 and $\scl = 1346$ \kms\ for A2152, the latter
being the highest dispersion in their sample of 25 ``dense peaks.''
Recently, Barmby \& Huchra (1998) have carried out a detailed analysis of
the Hercules supercluster kinematics using 414 galaxy velocities. 
These authors employed a four-component model, including the three
Abell clusters and a ``dispersed supercluster'' component.
They derived much lower velocity dispersions of $821^{+68}_{-55}$ \kms\ 
and $715^{+81}_{-61}$ \kms\ for A2147 and A2152, 
respectively, from 93 and 56 galaxy velocities.  

Because 122 galaxies in the Barmby \& Huchra sample are 
assigned to the dispersed component with a high velocity
dispersion of 1407 \kms, rather than to one of the individual
clusters, this approach may tend to bias the cluster dispersions low.
However, it is clearly more consistent in finding a mean redshift of
$\langle z \rangle=0.0432$ for A2152, much closer to that of the BCG,
whereas previous studies found means 10--15\% lower, near the mean
values for A2147 and A2151.  Moreover, the lower velocity dispersions are
more in line with the X-ray properties of these clusters, e.g., the
X-ray gas temperature in A2152 has been estimated from {\it Einstein} data
to be $k{T_X} = 2.1$~keV (White, Jones, \& Forman 1997).  We note, however,
that the latest tabulation by Struble \& Rood (1999) lists 
$\langle z \rangle=0.0398$ and $\scl = 1338$ \kms\ for A2152 from 
62 velocities.

The centroid of the X-ray emission associated with A2152 is not coincident
the BCG but is about 2\farcm1 east.  Its position is within 40\arcsec\ of
a pair of bright early-type galaxies which the Barmby \& Huchra redshift
catalogue reveals to be at $z=0.134$.  Each of these background galaxies
is intrinsically $\sim\,$70\% more luminous than the A2152 BCG
(using a relative K-correction of 0.11\,mag for ellipticals).
We return to this point in~\S\ref{sec:disc}.

\section{Observations}

A2152 was observed as part of the deep $R$-band Keck imaging study
by Blakeslee (1999) of the BCG halo regions in six nearby rich clusters.
Clusters were selected to have velocity dispersions
similar to that of the Coma cluster and to lie in the redshift range
$c{z}{\,=\,}$10,000--20,000 \kms.
These six were then observed because they were reachable
during a single night in the northern spring.  Based on this dataset,
Blakeslee \& Metzger (1999) reported follow-up spectroscopy of a
magnitude $m_R=20.86$ lensed arc located 27\arcsec\ from the
center of the cD galaxy in Abell 2124.  We have visually searched the
dataset for other bright lensing candidates, based on
proximity to the BCG, tangential orientation, and arc-like appearance.
The only obvious bright candidate was an arc-like object in A2152.
We obtained follow-up Keck spectroscopy for this object and the 
A2152 BCG/SRG pair.

\subsection{Imaging Data}
\label{ssec:imaging}

The imaging observations have already been described in detail by Blakeslee
(1999).  Here we summarize them and report new photometric results for the
A2152 arclet.  A2152 was imaged with the Low Resolution Imaging Spectrograph
(LRIS, Oke \etal\ 1995) on the Keck~II telescope under photometric
conditions.  The image scale was 0\farcs211~pix$^{-1}$, and the total
integration was 2000\,s in the $R$~band.  Reductions 
were carried out using the Vista package.  The seeing in the final stacked image
was an exceptional 0\farcs53.  The photometry was calibrated to the Cousins
$R$~band using Landolt (1992) standard stars; the photometric zero point is
accurate to better than 0.02~mag.

Figure~\ref{fig:img} displays the A2152 Keck LRIS image, a smaller portion
of which was shown by Blakeslee (1999).
The possible lensed galaxy (galaxy ``A'') is located 25\farcs2 from the
center of the BCG at a position angle PA$\,=\,$11\fdg4. 
The major axis of the BCG is  at $11^\circ\pm2^\circ$  at 
radii $\lta15\arcsec$, where the ellipticity (difference of the
axis ratio from unity) of the light distribution 
is $\epsilon_\ell{\,=\,}0.10$. Beyond this
radius the BCG isophotes have significant overlap with those of the SRG
and they are sufficiently round that the angle is difficult to determine.
The isophotes of the SRG show significant twisting;
the PA of its major axis swings from $170^\circ$ at $r\lta5\arcsec$ to
$\sim136^\circ$ by $r\approx25\arcsec$, while the ellipticity stays
fairly constant around $\epsilon_\ell{\,=\,}0.32\pm0.03$.

\figurenum{2}
\vbox{\begin{center}\leavevmode\hbox{%
\epsfxsize=8.8cm 
\epsffile{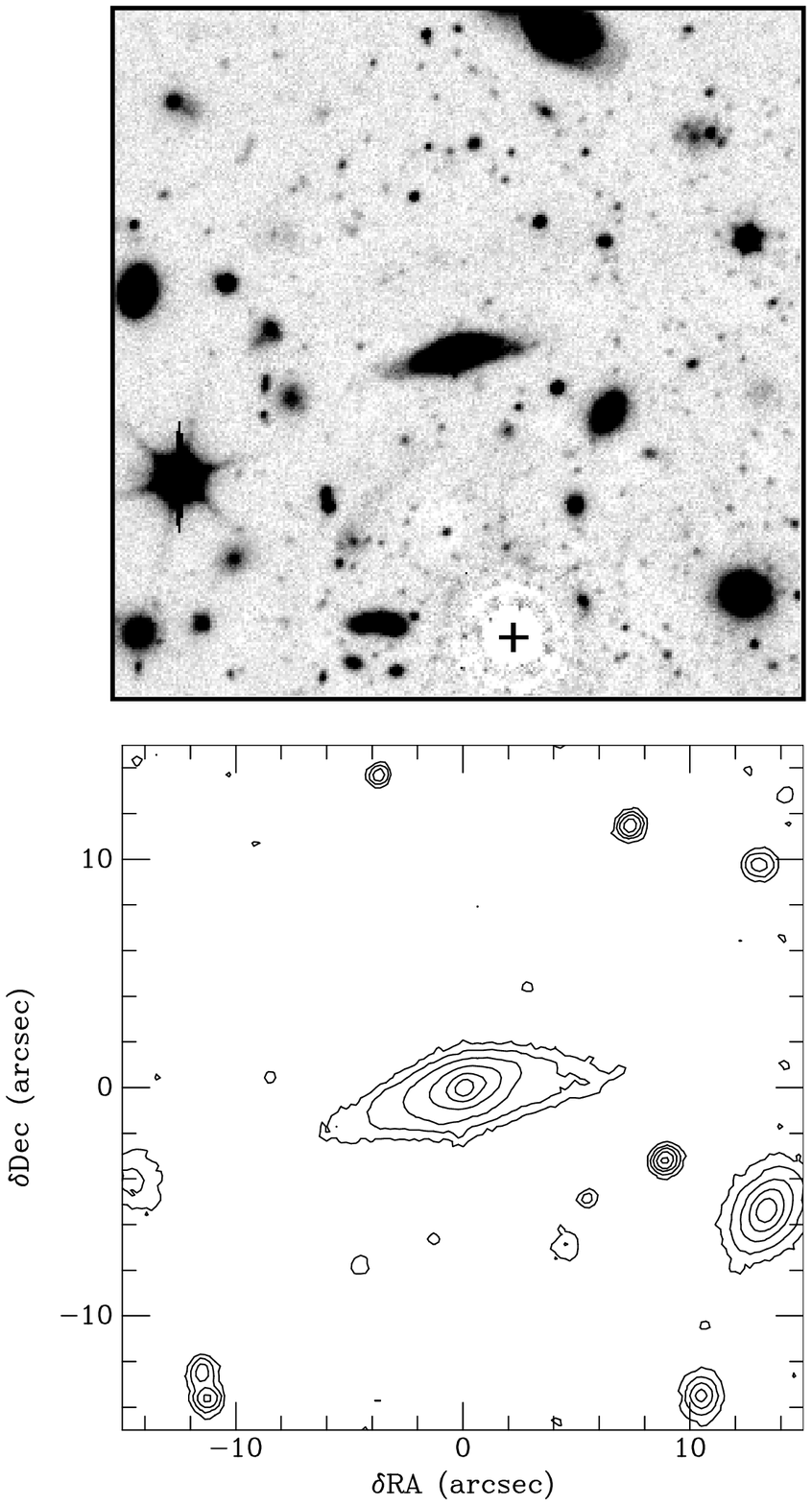}
%}\figcaption{\small
}\figcaption{\small
Two views of the possibly lensed galaxy in A2152 (galaxy A).
North is up and east is to the left.  The top view shows 
a $1\arcmin\times1\arcmin$ region of the $R$-band LRIS image
centered on the galaxy. The stretch is linear, and a model of the
BCG/SRG halo light has been subtracted.  The cross marks the center
of the subtracted BCG, which is 25\arcsec\ from A.
The lower panel zooms in by a factor of two, showing
an isophotal contour map of the $30\arcsec$ square
region around A.  The outermost contour is at
an extinction-corrected surface brightness of 
$24.8$ mag~arcsec$^{-2}$,
corresponding 2.0\% of the sky level in the image,
and the other contours are in steps of 1~mag.
The curvature of the object is apparent in both panels.
\label{fig:twoviews}}
\end{center}}

All galaxies with known redshifts (either from Barmby \&
Huchra or the present study) are labeled in Figure~\ref{fig:img}.  It is
somewhat curious that of 7 galaxies (including galaxy A) in the central
$\sim\,$6\arcmin\ field of a rich cluster, only 3 turn out to be cluster
members.  As mentioned above, the galaxy pair G1/G2 are in the background
with redshifts of $z=0.1335$ and $z=0.1353$, respectively. 
During our spectroscopic observation of the BCG/SRG (described in
\S\ref{sec:kinematics}), the LRIS slit picked up the magnitude $m_R=19.2$
galaxy labeled G3, which turns out to have $z=0.1326$.  Galaxy ``M'' is an
A2152 member with a velocity within $\sim 100$ \kms\ of the BCG (Barmby \&
Huchra).  The center of the X-ray emission in this field (Jones \& Forman
1999) is 20\arcsec\ from G2 and 50\arcsec\ from G1 and is also labeled in
Figure~\ref{fig:img}.

Figure~\ref{fig:twoviews} shows an enlarged view of galaxy A along with
an isophotal contour map.  The impression is that of a disk galaxy
whose disk has been warped into an arc-like shape
(arguably through lensing, as we discuss in \S\ref{sec:models}).
The object is visible in our image for a length of about 15\arcsec.
The measured full-width at half maximum along the minor axis
is 0\farcs67, which would be about
$\hbox{FWHM}\sim0\farcs4$ in the absence of atmospheric blurring.
This translates to a metric size $\lta 0.7$ \hkpc\ at the
redshift of this object (see below); apparently it has a very
compact nucleus.  The total magnitude of A,
corrected for 0.105~mag of $R$-band Galactic extinction
(\cite{SFD98}), is $m_R({\rm arc})=18.55\pm0.03$.
Our main measurement results are collected in Table~\ref{tab:res}.

\figurenum{3}
\begin{figure*}
\vbox{\begin{center}\leavevmode\hbox{%
\epsfxsize=13.0cm\epsffile{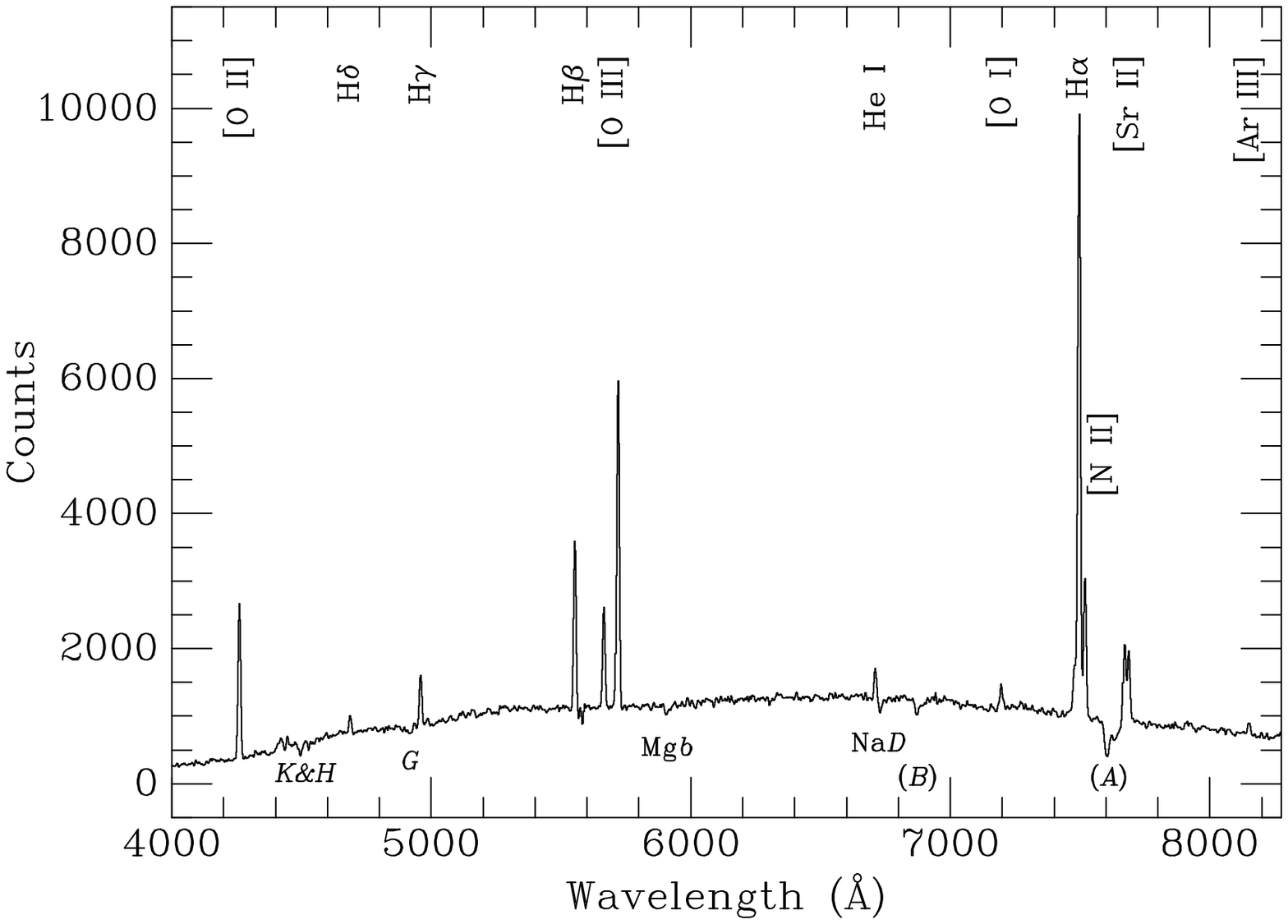}
}\end{center}
\smallskip
\figcaption{\small
Reduced 1-d spectrum of galaxy A.  The redshift is $z=0.1423$.
Various emission lines are identified above the spectrum
and some absorption features are labeled at bottom,
including the \ion{Ca}{2} $H$ and $K$ lines, $G$~band, Mg$\,b$,
and Na$\,D$.  The atmospheric $A$ and $B$ absorption bands are also
labeled; $A$-band absorption affects the [\ion{Sr}{2}] lines,
but not H$\alpha$.
\label{fig:arcspec}}}
\end{figure*}

\subsection{Spectroscopy of Galaxy~A}

The simplest and most important test of gravitational lensing is 
whether or not the supposed lensed object is actually in the background. 
Thus, we obtained one 900\,s spectrum of galaxy A with LRIS on the Keck~II
telescope on the night of 25~March 1999 (UT).  We used a 300 line grating
and a 1\arcsec\ wide slit oriented at $\hbox{PA}=113^\circ$, approximately
along the major axis of galaxy~A.  The seeing was poor, about 1\farcs8,
so the slit missed a good fraction of the light.  The dispersion was
2.44~\AA\,pix$^{-1}$, and the effective resolution was about 10~\AA\
(FWHM).  Halogen flats and arc-lamp exposures were taken for calibration.
The LRIS configuration and IRAF reductions were identical to those of
Blakeslee \& Metzger (1999), except that no flux standard spectrum was
obtained.

Figure~\ref{fig:arcspec} shows the reduced 1-d
spectrum of galaxy~A with the strongest emissions lines
labeled at top and some absorption lines labeled at the bottom.
The emission appeared to be confined to the central $\sim\,$4\arcsec\ 
of the galaxy, so this is all that was extracted.  The spectrum
is of very high signal-to-noise (it was not a~priori obvious,
in the absence of spiral structure or any spectral information,
that there whould be such strong features), and  
we determine the redshift to be $z=0.1423\pm0.0001$.
The object is therefore about 3.3 times more distant than the
intervening A2152 cluster, and the lensing hypothesis remains viable.
If unlensed, the absolute magnitude of A,
with a K-correction of $0.05$~mag, would be
% Omega=0/h=.7: dmod = 39.07;  1/.7: 38.99; 1/.5: 39.72
$M_R = -20.57$ 
% (for $H_0=70$ \kmsMpc and $\Omega{=}0$ or $M_R = -21.17$ 
% for $H_0=50$ \kmsMpc and $\Omega{=}1$);
(for $h=0.7$ and $\Omega{=}0$; $M_R = -21.22$ for $h=0.5$ and $\Omega{=}1$),
or a luminosity of about 0.48$\,L^*_R$ (e.g., Lin \etal\ 1996).

The H$\alpha$ emission line has an equivalent width of 
${\rm EW}_{{\rm H}_\alpha} = 117\pm6\,$\AA, corrected for
[\ion{N}{2}] emission. The H$\beta$ equivalent width 
is ${\rm EW}_{{\rm H}_\beta} = 28.2\pm2.0\,$\AA.
The [\ion{O}{2}] $\lambda3727$ and [\ion{O}{3}] $\lambda5007$ lines
have equivalent widths of $83.3\pm2.3\,$\AA and $54.1\pm2.0\,$\AA,
respectively.  Thus, the emission is quite strong, but not too 
anomalous in comparison to many late-type spirals or irregulars
(e.g., Kennicutt 1992).

Although we do not have a good absolute flux calibration of the spectrum,
we can roughly estimate the star formation rate (SFR) from the H$\alpha$ equivalent
width and the $R$-band luminosity.
We use the spectrophotometric synthesis results of Fukugita, Shimasaku, 
\& Ichikawa (1995) to calibrate the absolute flux in Cousins~$R$,
which has an effective wavelength of 6588\,\AA.
About 75\% of the $R$-band light comes from the region of emission within 
the galaxy, and from the measured H$\alpha$ equivalent width we obtain a flux
$f_{{\rm H}\alpha} \approx 7.38\times10^{-15}$ ergs~s$^{-1}$~cm$^{-2}$
and a luminosity of 
$L_{{\rm H}\alpha} \approx 1.84\times10^{41}$ $h^{-2}$\,ergs~s$^{-1}$
($\Omega=0$).  To estimate the internal extinction, we perform a makeshift
relative flux calibration using the Feige~110 spectrum from
Blakeslee \& Metzger (1999) and find an H$\alpha$-to-H$\beta$ flux ratio
of 4.25; assuming an unextincted value of 3.0
% for the Balmer decrement
and the extinction curve of Cardelli, Clayton \& Mathis (1989) gives
a correction at $H\alpha$ of a factor of 2.15.  This is 
close to the ``canonical'' correction of $\sim$1\,mag for spirals 
(e.g., Kennicutt 1992) but will yield a SFR 16\% lower.
Finally, using the updated calibration from Kennicutt (1998), 
we find $\hbox{SFR} \approx (3\pm1)\;h^{-2}\,\msun\,$yr$^{-1}$,
or $(6\pm2)\;\msun\,$yr$^{-1}$ for a typical $H_0$.
Though not qualifying as a ``starburst galaxy,'' the SFR is quite 
substantial for a 0.5$\,L^*$ galaxy.  The quoted errorbar is approximate
and reflects mainly the $\sim\,$30\% scatter quoted by Kennicutt. The
calibration based on [\ion{O}{2}] equivalent width yields nearly the same
SFR for this galaxy.

\subsection{Kinematics of the Central Galaxy}
\label{sec:kinematics}

With the intent of improving the constraints on the A2152
potential for the lensing models,
we obtained three 1100\,s spectra of the BCG/SRG pair
with LRIS on 16~June 1999.  We again used a 1\arcsec\ wide slit but
the 900/5500 grating, yielding a dispersion of 0.84 \AA\,pix$^{-1}$.
The slit was oriented at $\hbox{PA}=122^\circ$ in order to pass through the
centers of the two galaxies.  Again, halogen flats and arc lamps
were taken for calibration, and we also obtained spectra of the
K4~III velocity standard HD\,213947.  The spectra were processed,
wavelength calibrated, and rectified using standard IRAF routines.

Figure~\ref{fig:bcgspec} displays the central 1\farcs1 extracted
from the reduced 2-d spectra.  Although the spectra are clearly
very similar, the BCG exhibits a small amount of H$\beta$ emission
that is absent from the SRG.  Such optical line emission is quite
common among central cluster galaxies, and may be associated
with cluster ``cooling flows'' (e.g., Crawford \etal\ 1999).
This emission may therefore be evidence that the BCG is indeed 
at the center of the A2152 cluster potential and that the offset
of the Jones \& Forman (1999) X-ray position is due to contamination
from a background X-ray source.

The kinematical analysis of the data was performed within the MIDAS
package. First, the spectra were rebinned along the slit (spatial
direction) in order to achieve a minimum signal-to-noise (S/N) of 30
per \AA\/ at all radii. The S/N of the central pixels is higher. Then 
the continuum was removed by a fifth order polynomial fit. Finally we
determined the velocity dispersion and rotational velocity as a
function of radius with the Fourier Correlation Quotient (FCQ) method
(Bender 1990) allowing only for a Gaussian stellar velocity distribution.
When we allowed for non-Gaussian distributions and fitted for the 
higher-order moments, the results for the rotational velocity and
dispersion were unchanged within the errors but showed slightly
more scatter at the largest radii.
The detailed FCQ setup, such as continuum fit order, wavelength
extraction and error estimation was optimized with 
Monte Carlo simulations. 

\figurenum{4}
\vbox{\begin{center}\leavevmode\hbox{%
\epsfxsize=8.6cm
\epsffile{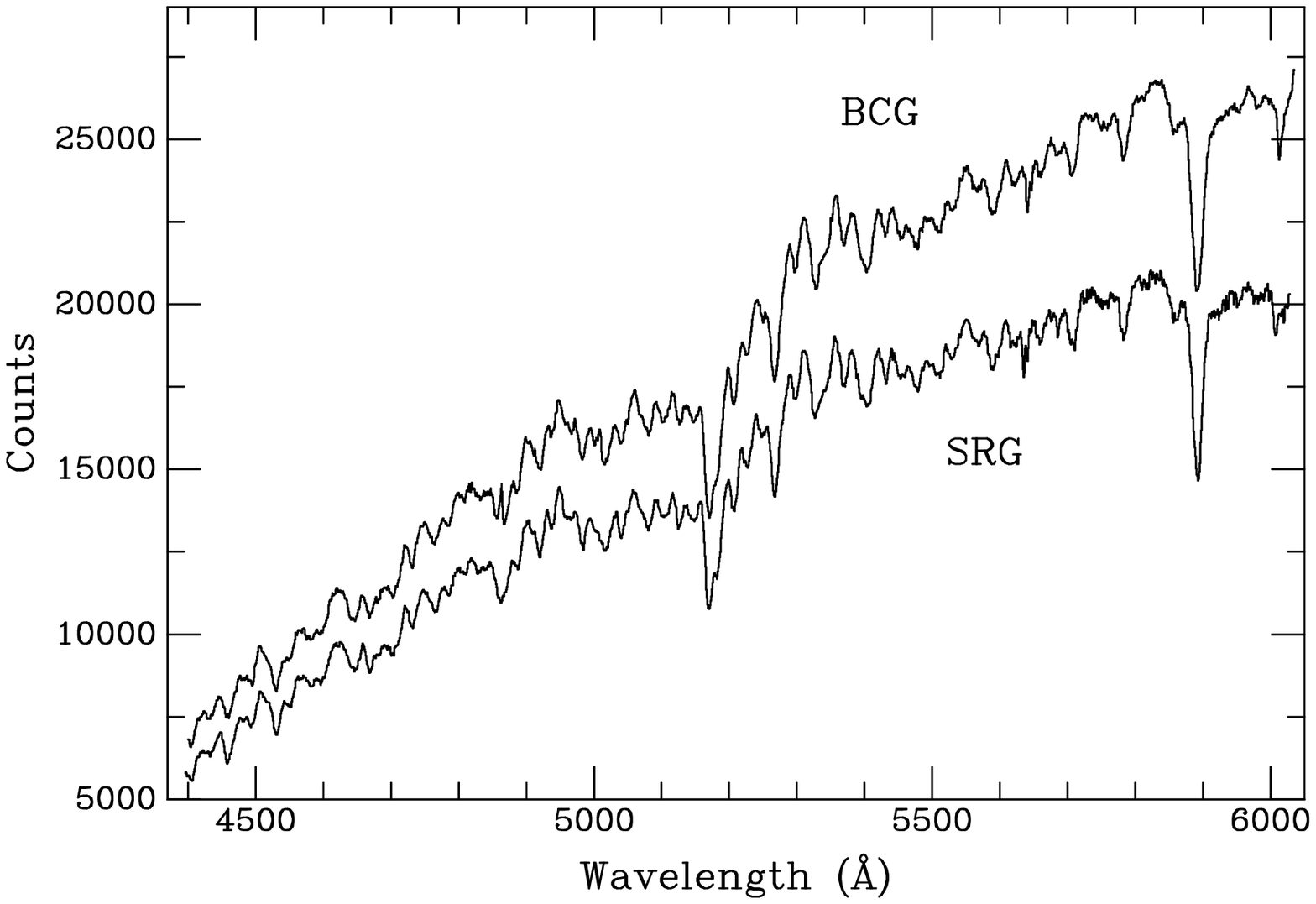}}
\figcaption{\small
The central 1\farcs1 extracted from the 2-d spectra of the 
BCG and SRG (observed simultaneously through the slit) are illustrated.
The spectra have been deredshifted for purposes of comparison.  
Some emission is visible within the H$\beta$ absorption at 
$\lambda{\,=\,}4861$ \AA\ in the spectrum of the BCG, but not in the SRG.
\label{fig:bcgspec}}
\end{center}}

Figure~\ref{fig:vprofs} shows the measured velocity and velocity 
dispersion profiles for the BCG and SRG. 
Note that the $122^\circ$ position angle is $\sim70^\circ$ away from
the major axis of the BCG, and so the profile is more appropriate
to the minor axis.  For the SRG, the PA is closer to the major axis.
Both galaxies show significant rotation, and both rotate in
the sense opposite to that of the combined BCG/SRG system.
Although the velocity dispersions decline outward from the centers of
the galaxies, the dispersion profile for the SRG appears to flatten 
at a value of $\sigma\sim230$ \kms\ outside the central 3\arcsec.
Likewise, the BCG's velocity dispersion profile does not show 
a significant decrease for $r\gta7\arcsec$.
As seen in the top panel of Figure~\ref{fig:vprofs}, 
the center of rotation in the BCG is offset by about
0\farcs2 (1 LRIS pixel) from the luminosity center; this is likely
due to a slight offset of the galaxy center within the slit.
The mean velocities, rotational velocities, and central dispersions
are tabulated along with our other results in Table~\ref{tab:res}.

We have estimated the total (1-sigma) errors 
on the mean velocities to be $\pm25$\,\kms,
but this is dominated by zero-point calibration errors (flexure, illumination
of slit, etc.).  However, the relative velocity of the BCG/SRG pair 
is better determined at $\Delta v = 331\pm7\,$\kms.  This accords well
with previous measurements of $\Delta v = 327\pm100$ \kms\ from 
Tarenghi \etal\ (1979) and  $\Delta v = 341\pm50$ \kms\ from 
Wegner \etal\ (1999); less well with $\Delta v = 502\pm21$ \kms\ from 
Davoust \& Consid\`ere (1995). (The velocities used by Barmby \& Huchra
for these two galaxies come from Tarenghi \etal).

\figurenum{5}
\vbox{\begin{center}\leavevmode\hbox{%
\epsfxsize=8.6cm
\epsffile{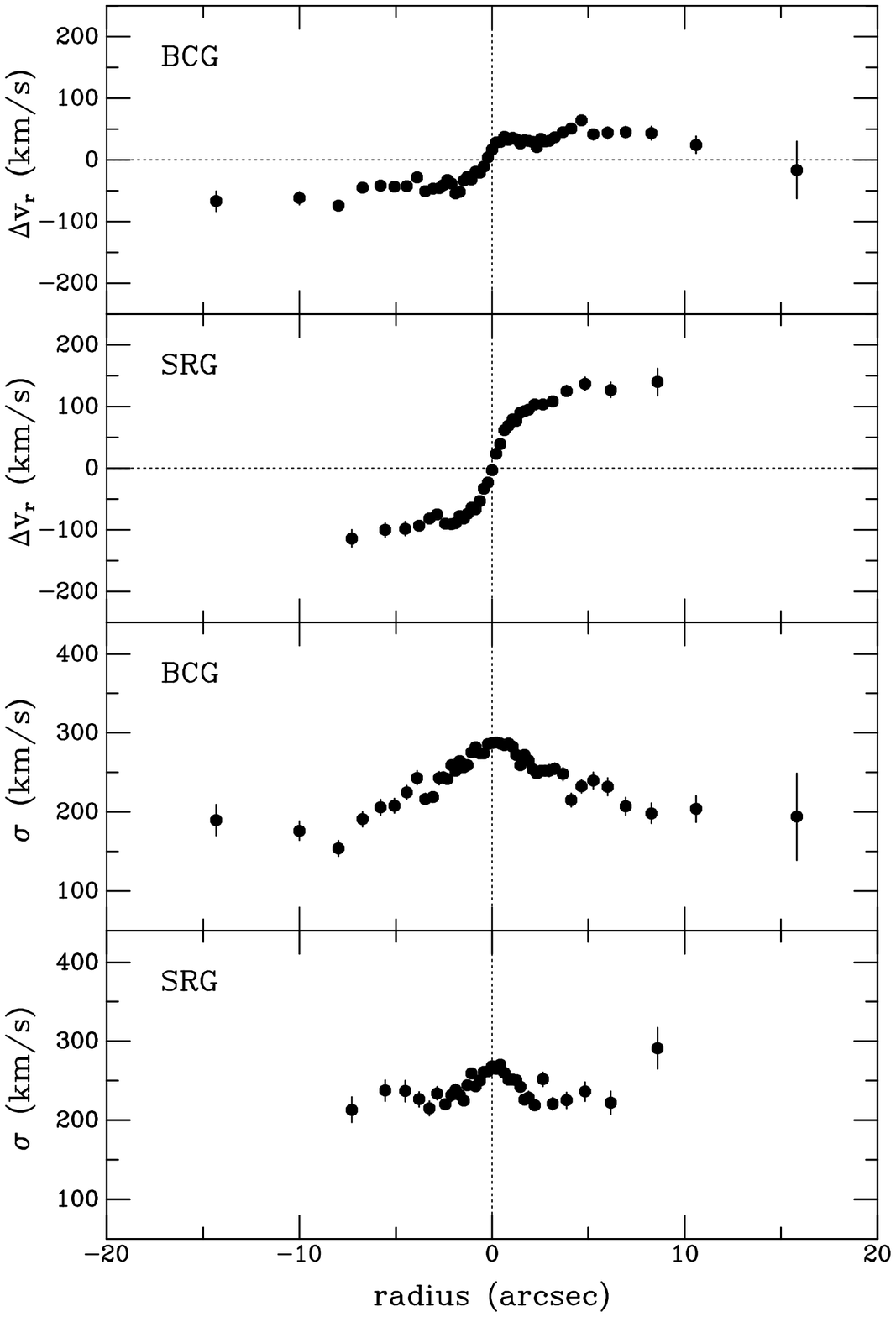}}
\figcaption{\small
Velocity and dispersion profiles for the BCG and SRG (labeled).
Note that the radial coordinate defined here increases to the
southeast along the slit, i.e., the BCG is in the direction of
positive radius with respect to the SRG.  Thus, the point
in the SRG's dispersion profile at $(r,\sigma) = (8\farcs6,291)$
may be beginning to show the effects of the BCG halo light.
\label{fig:vprofs}}
\end{center}}

Central velocity dispersions have been measured previously for the BCG by
Oegerle \& Hoessel (1991), who found $\sigma_0 = 295\pm25$ \kms, and by
Wegner \etal\ (1999), who reported $\sigma_0 = 280\pm26$ \kms; both results
are in agreement with our much more precise value of $\sigma_0 = 295\pm7$
\kms.  For the SRG, we find a central dispersion of $\sigma_0 = 271\pm8$
\kms, while Wegner \etal\ found $\sigma_0 = 313\pm34$ \kms. Their
significantly higher dispersion is likely due to the strong rotation in
this galaxy, which these authors did not account for when extracting the
spectrum of the central several arcseconds of the galaxy.  Finally, our BCG
mean velocity is $\sim2.5\,\sigma$ discrepant with the Wegner \etal\ value,
which is higher by 95\,\kms, but it agrees within the errors with the
values from Tarenghi \etal\ (1979), Oegerle \& Hoessel (1991), and Davoust
\& Consid\`ere (1995).  No previous measurements of the rotation or
dispersion profile in either galaxy could be found.

Finally, as noted in \S\ref{ssec:imaging}, the LRIS slit picked up
the magnitude $m_R = 19.24$ galaxy labeled G3 in Figure~\ref{fig:img}.
We find it to be in the background with a velocity $c{z} = 39{,}738$ \kms.
This is just $\sim290$ \kms\ less than the velocity of G1 and $\sim830$ \kms\
less than that of G2.

\section{Is it Lensed?}
\label{sec:models}

To model A2152 as a lens, we use a potential of the
form (Blandford \& Kochanek 1987)
$$
\psi(r^\prime) = 4 \pi {\left( {\sigma_{1D} \over c } \right)}^2
{{D_{LS}} \over {D_S}} \left[ {\left( 1 + { {r^\prime}^2 \over r_c^2 }
\right) }^{1 \over 2} - 1 \right] ,
$$
where $\sigma_{1D}$ is the line-of-sight velocity dispersion in the
limit $r^\prime \gg r_c$ for the spherical case, $D_S$ and $D_{LS}$ 
are angular size distances to the source and from the lens to 
the source, respectively, and $r_c$ defines a softening radius.
An ellipticity parameter $\epsilon_p$ can be introduced through
${r^\prime}^2 = (1{-}\epsilon_p) x^2 + (1{+}\epsilon_p) y^2$,
where $(x,y)$ are coordinates aligned with the major and minor axes
of the potential.  This represents a softened isothermal sphere for
$\epsilon_p = 0$ and provides a fairly good representation 
of the dark halo potentials in the cores of clusters 
(e.g., Mellier \etal\ 1993; Tyson \etal\ 1998). 
We use the more limited relation with a fixed power law because
there are too few lensing constraints in this system to
differentiate various potential profiles, and we have used 
only circularly symmetric potentials
as there is little constraint on the ellipticity.

We explored four different simple models for the 
lensing potential.  We included three terms in each of these
models, corresponding to the smooth halo mass in the cluster 
and the masses associated with the BCG and SRG.  
For the latter two components, we use analytic potentials to approximate the
velocity dispersions measured in \S\ref{sec:kinematics} with a small core.
In our model~1, the
center of the cluster potential was fixed at the cD center with
an asymptotic dispersion of $\sigma_{\rm cl}=715$ \kms\ as measured by
Barmby \& Huchra (1998) and a softened core radius of 10\arcsec.
Model 2 also fixes the cluster potential at the BCG, but uses the higher
dispersion estimate of 1340 \kms\ (Struble \& Rood 1999). 
Models 3 and 4 are the same as models
1 and 2, but fix the cluster potential center at the X-ray center (while
of course keeping the BCG and SRG model components fixed).  Within each
model, we also explored a range of core softening radii.

Our primary constraints for the lens models are the radius, size, and
orientation of galaxy~A.  The modest distortion and lack of a counterarc
suggests a source position that lies outside the tangential and radial
caustics (e.g., Grossman \& Narayan 1988) and whose image lies outside
the critical curve.  These constraints quickly rule out model~2 in
most forms: with a small core radius, the critical curve for $z=0.142$
lies at around 35 arcsec, well outside the galaxy radius.
This would imply multiple images and large distortions for galaxy A.
Even with a large core radius, the effective radial magnification
becomes large and can be ruled out by the lens appearance.  Model 3
does not produce significant magnification or distortion for the galaxy;
thus under such a model, the arc-like shape would need to be intrinsic.
Model 4 produces shear and magnification at the position of galaxy A,
but the shear direction is not directed tangential to the BCG, as
observed in galaxy A.

Model 1, however, produces modest magnification and distortion of a
$z=0.142$ source with a shear aligned tangentially to the BCG/cluster.
A family of models with a range of core radii and asymptotic
dispersions for the primary lens (cluster dark matter) can 
reproduce the characteristics of the arc.  These models are
degenerate in $\sigma_{1D}$ and $r_c$ for determining the mass
enclosed within the arc radius.
Model 1 fits within this family of models.
% with either a 10\arcsec\ or 40\arcsec\ core.
Figure~\ref{fig:lensmod} shows the predicted magnifications from 
model~1 with a 10\arcsec\ core as a function of observed source position.
The position of galaxy A implies a magnification of $\sim 1.9$,
dropping to $\sim 1.8$ for a model with a 40\arcsec\ soft core.
Given this model, we can ask what the intrinsic source shape
of galaxy A would be; Figure~\ref{fig:lensmap} shows contours of the
observed galaxy surface brightness along with the same contours 
mapped back to the source plane via the lens equation.  The implied
shape is that of a fairly symmetric galaxy with a disk and dominant
bulge, with no indication of a warp.

While we cannot make quantitative estimates of cluster mass parameters with
a single lensed galaxy, we find that a simple model constructed from a
low-dispersion cluster centered at the BCG can produce the shear needed to
explain the curved shape of the background galaxy from a more symmetric
source.  Indeed, galaxy A must be distorted in some fashion by the
foreground cluster, and we can rule out two of our other three plausible
models based on the implied lensing geometry.  We cannot, however, rule out
a low-dispersion cluster mass centered on the X-ray center from lensing
alone, although this would mean galaxy~A must be intrinsically warped.
Another galaxy with an unusual shape, elongated {\it radially} from the BCG
center, is visible about 11\arcsec\ to the east of the BCG (see the upper
panel of Fig.~\ref{fig:twoviews}) and lies near some critical radii of
model~1.  If in the background, this galaxy may provide much better
constraints on the shape of the inner potential, however the redshift is
unknown and we do not attempt to model it here.

\figurenum{6}
\vbox{\centering\leavevmode\hbox{%
\epsfxsize=8.4cm
\epsffile{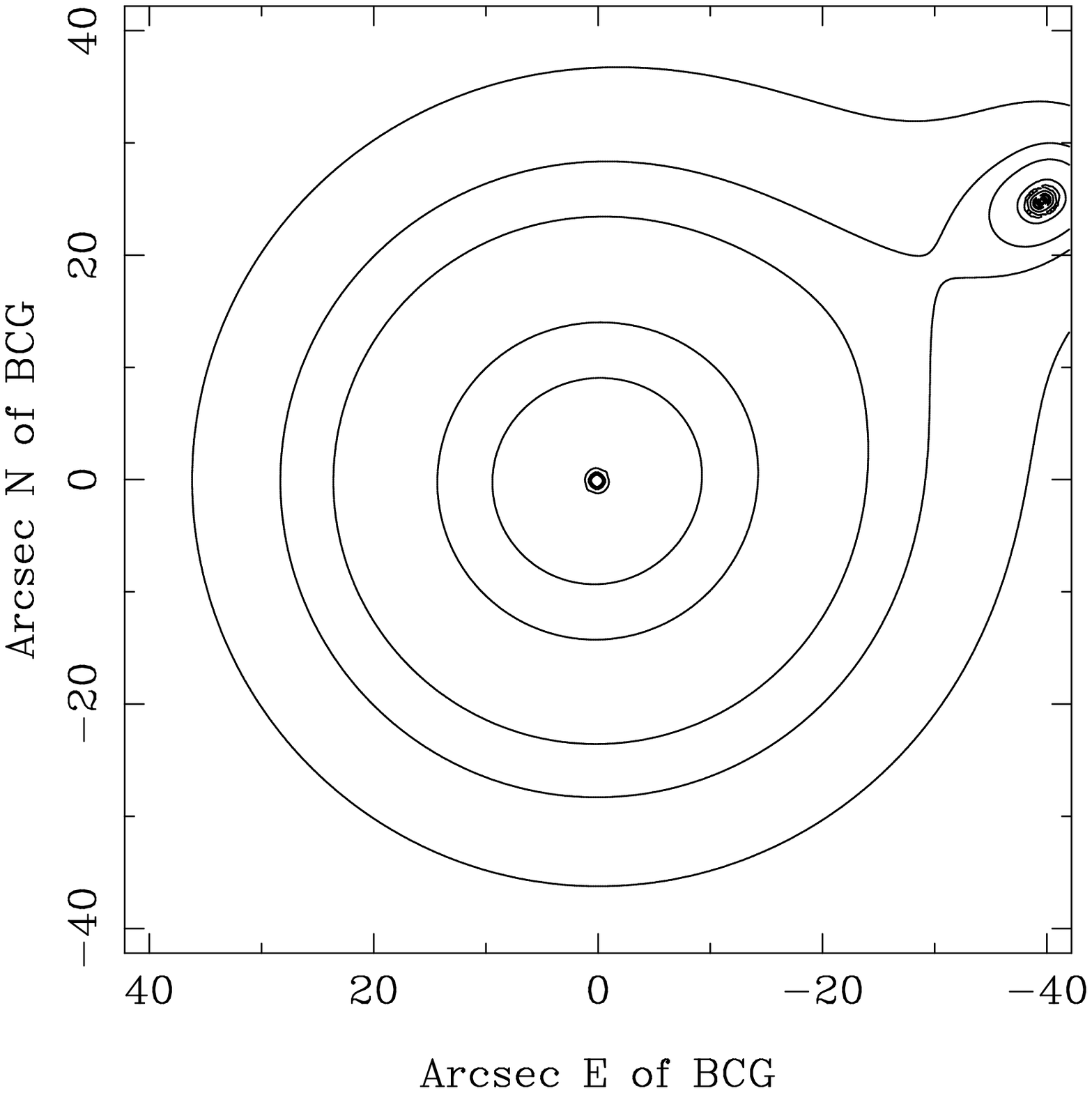}} 
\figcaption{\small Magnification of a
background source at $z=0.142$ produced by the simple lens model.  
The model shown has $\sigma_{\rm cl}=715$ \kms\ with a 10 arcsec soft
core, $\sigma_{\rm BCG}=250$ \kms, and $\sigma_{\rm SRG}=240$ \kms.
Starting from the outside, contours indicate magnifications of 1.5,
1.73, 2, 4, and 16.  The critical curve is near the magnification=16
contour, and a second critical curve is present close to the center.
\label{fig:lensmod}}
}

\figurenum{7}
\vbox{\begin{center}\leavevmode\hbox{%
\epsfxsize=8.6cm
\epsffile{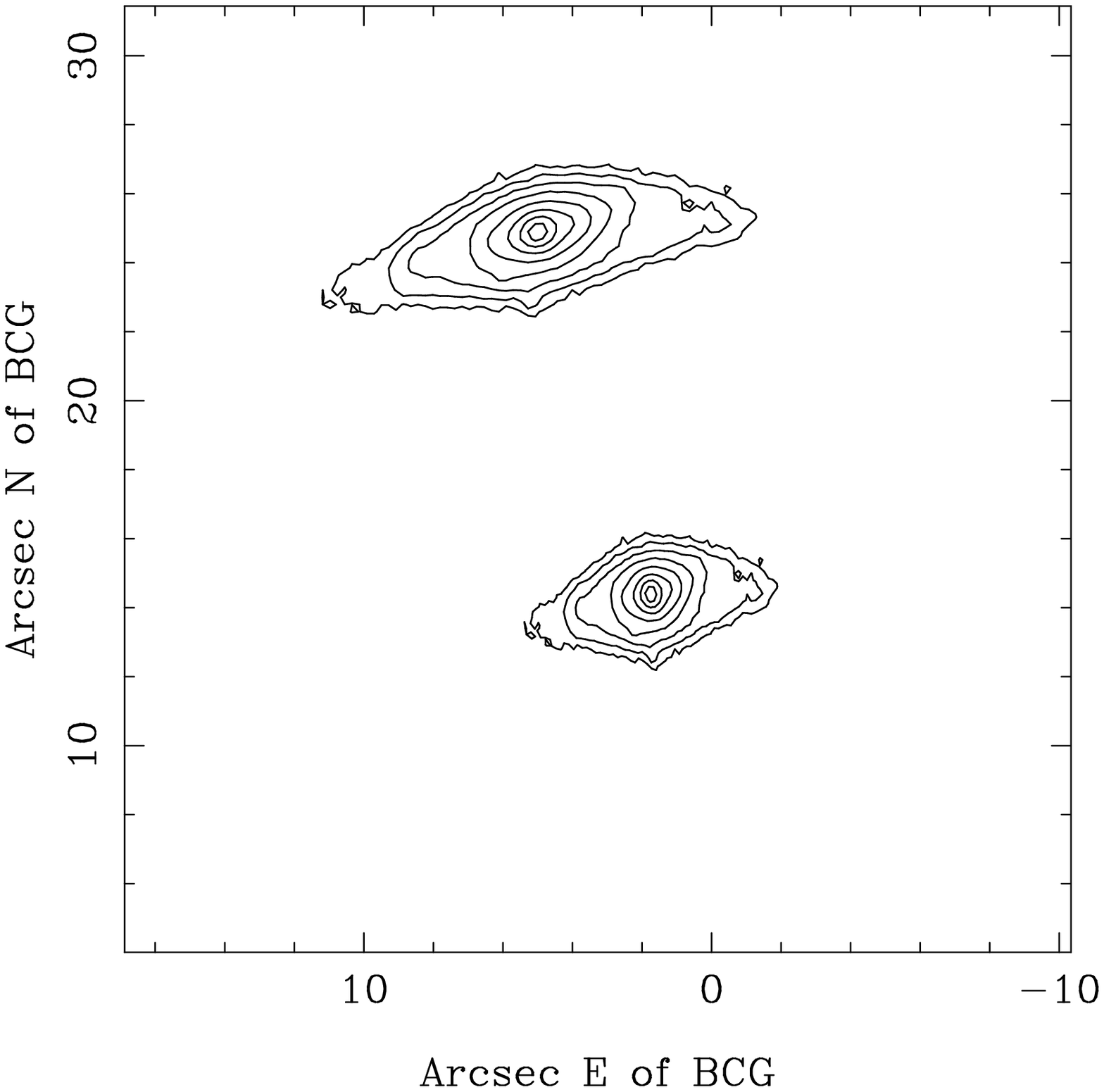}} 
\figcaption{\small Contours of the observed
galaxy and the inferred source contours from the simple lens model.
The upper image shows contours of the observed surface brightness of 
the lensed galaxy A, spaced by factors of two. The lower image shows
the implied source position and the same surface brightness contours
on the source plane, i.e. how the galaxy would appear if the foreground
cluster were absent.
%%% add this note...
The innermost contours, less than 1\arcsec\ in size, are rounded in
the observed image from the seeing; therefore, these contours become
artificially elongated in the reconstructed source image.
The coordinates are centered on the BCG.
\label{fig:lensmap}}
\end{center}}

\section{Summary and Discussion}
\label{sec:disc}

Deep $R$-band Keck imaging taken under excellent seeing conditions has 
revealed a curved object 25\farcs2 (15 \hkpc) from the center of the
A2152 BCG, along the major axis.
The reddening-corrected
$R$~magnitude of this ``arc'' is $m_R=18.55\pm0.03$, and follow-up 
spectroscopy shows that it lies at a redshift $z=0.1423\pm0.0001$. 
Its position, orientation, and redshift all
suggest a lensing explanation for the curiously bowed structure. 
The object has numerous strong emission lines, and the equivalent width of
H$\alpha$ is $117\pm6$ \AA, from which we infer a star formation rate
of $(3\pm1)\;h^{-2}\,\msun\,$yr$^{-1}$.

We have presented spatially resolved spectroscopy of the A2152 BCG and SRG,
which are separated by 47\arcsec\ and are together catalogued as UGC\,10187. 
The relative velocity of the BCG and SRG is $331\pm7$ \kms, and
both galaxies show significant rotation.
The rotational velocities along the observed position angle are
$\sim50$ \kms\ and $\sim120$ \kms\ for the BCG and SRG, respectively.
Our velocity dispersion profile for the BCG extends to a radius
of $\sim16\arcsec$ (9.5\,\hMpc), about 60\% of the way to the 
radial position of the lensed galaxy.

We have modeled the lensing potential of A2152 using constraints
from the measured cluster velocity dispersion of 715 \kms\
and our own stellar velocity dispersion measurements for the BCG and SRG.
The models cannot produce a significant lensing effect
on galaxy A if the  A2152 potential center is taken to coincide
with the peak of the overall X-ray emission.
However, if the potential center is taken to be the A2152 BCG, then
the models {\it can} reproduce the observed position and apparent 
bowed shape of the galaxy, 
and the implied magnification is a factor of about 1.9.  Conversely,
we can say that if the object is significantly lensed, then the 2\farcm1
offset of the X-ray centroid from the BCG is due to contamination from
another source of X-ray emission.  
We note that both the cluster velocity
dispersion and the core softening radius in our lensing model 
is degenerate with the source position. 
However, the higher values that have been reported for the dispersion
(see \S\ref{sec:clusprops}), if centered on the BCG, produce lens
models with critical curves that lie outside of the lensed galaxy,
thus multiple images would be expected of any background galaxy at the
position and redshift of A.  We can therefore rule out this case from
a lensing argument; even a large core radius model would produce significantly
more distortion in a $z=0.142$ galaxy.

The available redshift data provide evidence for a background association of
early-type galaxies at $z\approx0.134$, projected near the center of A2152.
Two of the galaxies, which we call G1/G2, have redshifts from the
Barmby \& Huchra (1998) catalogue. They are separated by 30\arcsec\ 
and are each about 70\% more luminous than the A2152 BCG.
The third galaxy, called G3, is 35\arcsec\ southeast of the A2152 BCG
and was picked up by the LRIS slit during our observations.  It is 116\arcsec\
from G1 and 132\arcsec\ from G2, or metric distances of 188-215 \hkpc.
The velocity range for these three galaxies is $\sim830$ \kms.
While three galaxies do not a cluster make, the luminosities, velocities,
and early-type morphologies of the galaxies are highly suggestive. 
Moreover, there is a strong concentration of small galaxies of unknown
redshift centered near the bright G1/G2 pair.
It is highly likely that some of the X-ray emission, which is
centered just 20\arcsec\ from G2, could be due to a grouping of galaxies
associated with G1/G2.
Therefore, the center of the true A2152 X-ray emission, and the center of
the A2152 cluster potential, might well coincide with the A2152 BCG, as
required by our lensing models.  We note that the centers of the outermost
X-ray contours in the Jones \& Forman map of this field drift away from
G1/G2 and closer to the A2152 BCG; the optical H$\beta$ emission we find in
the BCG spectrum may be additional circumstantial evidence for central
location in the cluster potential.  However, a high-resolution X-ray map is
needed to confirm this hypothesis.

If there is a background cluster, galaxy~A may well be a member,
in which case it would be falling away from us through the cluster with a
velocity of $\sim2500$ \kms.  Simulations show that the tidal effects 
of clusters on infalling galaxies
can warp spiral disks into arc-like shapes (Moore \etal\ 1999a)
and even form giant arcs of tidal debris (Calc{\'a}neo-Rold{\'a}n \etal\ 2000).
Thus, tidal interaction with a local cluster potential, rather than 
lensing by the foreground A2152 potential, could be
responsible for the distorted appearance of the galaxy, as well as for the
observed strong star formation.  Its location and orientation with respect
to the A2152 BCG would then be merely fortuitous.
However, the galaxy in question would have to be
caught during its initial pass through the cluster, since gas in cluster
galaxies tends to be lost on relatively short time-scales, truncating the star
formation (e.g., Balogh \etal\ 1998; Kodama \& Bower 2000).  
%%%%
The probability that galaxy A is a member of a cluster at
$z=0.13$ depends of course on the velocity dispersion of the
% (still hypothetical) 
cluster.  A deep redshift survey of this field is needed to discern whether
or not there is a very substantial background cluster (as suggested by the
luminosities of G1/G2 and the position of the X-ray center in this field),
and how likely galaxy A's membership is.  Such a redshift survey would need
to go at least $\sim2$~mag fainter than the Barmby \& Huchra (1998) limit
in order to pick up the multitude of small galaxies that lie between the
A2152 BCG/SRG pair and the G1/G2 background pair.

Although superficially A2152 might have seemed an 
unremarkable low-redshift cluster (leaving aside the
unlikely velocity dispersions in excess 1300 \kms\ found in the literature),
our photometric and spectroscopic observations of its central field have
raised a host of intriguing new questions
about this cluster and the arc-shaped galaxy projected near its BCG.
How much mass is there along this line of sight and how is it
distributed?  Is the A2152 BCG actually at the center of the cluster
potential and the observed offset with respect to the X-ray center due
to the superposed emission of a background cluster at $z=0.13$?  If
so, is the distorted galaxy a member which is being distorted by the
hypothetical background cluster, or is it a field galaxy being lensed
by A2152?
Along with high-resolution X-ray observations and a deep redshift survey,
a weak lensing analysis of this field would help in unraveling the
projected mass distribution and provide insight into these questions.

We believe that the weight of the evidence favors the view that the
A2152 BCG is at the center of its cluster's potential and is
significantly lensing galaxy A.  Further lensed objects, if present
near critical curves, could provide useful future constraints on the
central structure of rich clusters.  For now, however, a complete
understanding of this system and definitive answers to the questions
raised by this study must await an infusion of new data.

\acknowledgments 
We thank Michael Balogh, Alastair Edge, Ben Moore, Ian Smail, and Graham
Smith for helpful conversations.  This work made use of the NASA/IPAC
Extragalactic Database (NED), operated by the Jet Propulsion Laboratory at
Caltech under contract with the National Aeronautics and Space
Administration.  It also made use of Starlink computer facilities.  We are
grateful to the team of scientists and engineers responsible for producing
the Low Resolution Imaging Spectrograph.  J.P.B. and P.C. thank the Sherman
Fairchild Foundation for support while at Caltech.  M.R.M's research was
supported by Caltech.  J.P.B. and H.K. were supported at the University of
Durham by a PPARC rolling grant in Extragalactic Astronomy and Cosmology.

\par\medskip\noindent{\it
Note added in proof\,---\,New multi-color data confirm the presence of
a rich background cluster around the $z{\,=\,}0.13$ G1/G2 galaxy pair;
analysis of these data will be presented in a forthcoming paper.}

%\clearpage

\begin{deluxetable}{lcrr}\tablenum{1}
\centering
\small 
\tablecaption{ Summary of Measurements}\label{tab:res}
\tablewidth{240pt} \tabcolsep=0.5cm
\tablehead{ 
\colhead{Object} & \colhead{quantity} & \colhead{value} & \colhead{$\;\pm$}
}\startdata
Arclet & $m_R$\tablenotemark{(a)}        & 18.55 & 0.03  \\
G3     & $m_R$\tablenotemark{(a)}       & 19.24 & 0.05  \\
Arclet & {\sc fwhm}\tablenotemark{(b)}   & 0\farcs42 & 0\farcs08  \\
BCG    & $\epsilon_\ell$\tablenotemark{(c)}    & 0.10 &  0.01  \\
SRG    & $\epsilon_\ell$\tablenotemark{(c)}    & 0.33 &  0.02  \\
BCG    & PA\tablenotemark{(d)}      & 10\fdg7 &  0\fdg7  \\
SRG    & PA\tablenotemark{(d)}      & 161\fdg4 &  2\fdg0  \\
Arclet & $c{z}$\tablenotemark{(e)}      & 42665 &  40  \\
BCG    & $c{z}$\tablenotemark{(e)}      & 13188 &  25  \\
SRG    & $c{z}$\tablenotemark{(e)}      & 13519 &  25  \\
G3     & $c{z}$\tablenotemark{(e)}      & 39738 &  25  \\
BCG    & $v_m$\tablenotemark{(f)}  &  49  &   5  \\
SRG    & $v_m$\tablenotemark{(f)}  & 117  &   7  \\
BCG    & $\sigma_0$\tablenotemark{(g)}  &   295 &   7  \\
SRG    & $\sigma_0$\tablenotemark{(g)}  &   271 &   8  \\
\enddata
\vspace{-5pt}
\tablenotetext{(a)}{\footnotesize $R$~magnitude, corrected for Galactic extinction}
\tablenotetext{(b)}{\footnotesize seeing-corrected {\sc FWHM} along the minor axis of arclet}
\tablenotetext{(c)}{\footnotesize ellipticity of the light distribution, at $r{\,=\,}10\arcsec$}
\tablenotetext{(d)}{\footnotesize angle of major axis, east from north, at $r{\,=\,}10\arcsec$}
\tablenotetext{(e)}{\footnotesize heliocentric redshift (\kms)}
\tablenotetext{(f)}{\footnotesize rotational velocity (\kms) at $r{\,\gta\,}5\arcsec$ and PA${\,=\,}122^\circ$}
\tablenotetext{(g)}{\footnotesize central velocity dispersion (\kms)}
\end{deluxetable}

%%% uncomment the \epsfile command to include the postscript for figure 1
\figurenum{1}
\begin{figure*}
\vbox{\begin{center}\leavevmode\hbox{%
\epsfxsize=17.0cm
%%\epsffile{Figs/fig1.ps}
\centerline{\huge fig1.gif }
}\end{center}
\medskip
\caption{\small
An A2152 finding chart.
The image was taken with Keck/LRIS through an $R$ filter.
The field is $5\farcm6{\,\times\,}6\farcm0$ in size, and the
bright galaxy at center is the A2152 BCG.  Several other galaxies
are labeled: the second-ranked cluster galaxy (SRG);
the arclet at $z=0.142$ (A); three early-type galaxies all
at $z{\,\approx\,}0.133$ (G1, G2, G3); and the only other galaxy
in this field with a known redshift (M).  The redshifts of the
BCG, SRG, and galaxy ``M'' are all $z{\,\approx\,}0.044$.  
For reference, the arclet is 25\arcsec\ from the center of the BCG,
and the BCG/SRG pair are separated by 47\arcsec.
The large ``X'' near the bright pair of background galaxies G1/G2
marks the location of the centroid of the X-ray emission, normally
attributed to A2152 (Jones \& Forman 1999).
\label{fig:img}}}
\end{figure*}


\begin{thebibliography}{}

\bibitem[Abell \etal\ 1961]{Abell61} Abell, G.\ O. 1961, \aj, 66, 607

\bibitem[Abell \etal\ 1989]{ACO89}
Abell, G.\ O., Corwin H.\ G.\ \& Olowin, R.\ P. 1989, \apjs, 70,~1

\bibitem[Allen, Fabian and Kneib (1996)]{1996MNRAS.279..615A}
Allen, S. W., Fabian, A. C. and Kneib, J. P. 1996, \mnras, 279, 615 

\bibitem[Balogh, et al. (1998)]{1998ApJ...504L..75B} 
Balogh, M. L., Schade, D., Morris, S. L., Yee, H. K. C., Carlberg, R. G. \&
Ellingson, E. 1998, \apjl, 504, L75 

\bibitem[Barmby \& Huchra 1998]{BH98}
Barmby, P. \& Huchra, J. P. 1998, \apj, 115, 6

\bibitem[Bender (1990)]{1990A&A...229..441B} Bender, R. 1990, \aap, 229, 441 

\bibitem[Bird \etal\ 1995]{BDB95}
Bird, C. M., Davis, D. S., \& Beers, T. C. 1995, \aj, 109, 920

\bibitem[Blakeslee 1999]{Blak99} 
Blakeslee, J. P. 1999, \aj, 118, 1506

\bibitem[Blakeslee \& Metzger 1999]{BM99}
Blakeslee, J. P. \& Metzger, M. R. 1999, \apj, 513, 592

\bibitem[Blandford \& Kochanek 1987]{BK87}
Blandford, R. D. \& Kochanek, C. S. 1987, \apj, 321, 658

\bibitem[Calc{\'a}neo-Rold{\'a}n, et al. (2000)]{2000MNRAS.314..324C} 
Calc{\'a}neo-Rold{\'a}n, C., Moore, B., Bland-Hawthorn, J., Malin, D. \&
Sadler, E. M. 2000, \mnras, 314, 324 

\bibitem[Campusano \etal\ 1998]{CKH98}
Campusano, L. D., Kneib, J.-P., \& Hardy, E. 1998, \apjl, 496, L79

\bibitem[Cardelli, Clayton and Mathis (1989)]{1989ApJ...345..245C} 
Cardelli, J. A., Clayton, G. C. \& Mathis, J. S. 1989, \apj, 345, 245 

\bibitem[Crawford, et al. (1999)]{1999MNRAS.306..857C} 
Crawford, C. S., Allen, S. W., Ebeling, H., Edge, A. C. \& Fabian, A. C.
1999, \mnras, 306, 857 

\bibitem[Cypriano \etal\ (2000)]{Cyp2000}
Cypriano, E. S., Sodr\'e, L., Campusano, L. E., Kneib, J.-P., 
Giovanelli, R., Haynes, M. P., Dale, D. A., \& Hardy, E. 2000, \aj,
submitted (astro-ph/0005200)

\bibitem[Davoust and Considere (1995)]{1995A&AS..110...19D} 
Davoust, E. \& Consid\`ere, S. 1995, \aaps, 110, 19 

\bibitem[Dubinski and Carlberg (1991)]{1991ApJ...378..496D} 
Dubinski, J. and Carlberg, R. G. 1991, \apj, 378, 496 

% \bibitem[Ebeling \etal\ 1996]{Eb96}
% Ebeling , H., Voges, W., Bohringer, H., Edge, A. C., Huchra, J. P.,
% Briel, U. G. 1996, \mnras, 281, 799

\bibitem[FSI1995]{FSI95}
Fukugita, M., Shimasaku, K., \& Ichikawa T. 1995, \pasp, 107, 945

\bibitem[Grossman \& Narayan 1988]{GN88}
Grossman, S. A. \& Narayan R. 1988, \apj, 324, L37

% \bibitem[Kennicut 1992a]{K92supp} Kennicutt, R. C. 1992a, \apjs, 79, 255

\bibitem[Jones \& Forman 1999]{jf99}Jones, C. \& Forman, W. 1999, \apj, 511, 65

\bibitem[Kennicut 1992b]{K92apj} Kennicutt, R. C. 1992, \apj, 388, 310

\bibitem[Kennicut 1998]{K98} Kennicutt, R. C. 1998, \araa, 36, 189

\bibitem[Kodama \& Bower]{KB2000}
Kodama, T. \& Bower, R. G. 2000, \mnras, submitted (astro-ph/0005397)

\bibitem[Landolt 1992]{arlo} Landolt, A. U. 1992, \aj, 104, 340

\bibitem[LeF\`evre \etal\ (1994)]{LHAGL94}
LeF\`evre, O., Hammer, F., Angonin, M.\,C., Gioia, I.\,M., \& Luppino, G.\,A.
1994,~\apjl,~422,~L5

\bibitem[Lin \etal\ 1996]{lkslots96}
Lin, H., Kirshner, R. P., Shectman, S. A., Landy, S. D., Oemler, A.,
Tucker, D. L., \& Schechter, P. L., 1996, \apj, 464, 60

\bibitem[Mellier \etal\ 1993]{MFK93}
Mellier, Y., Fort, B., \& Kneib, J.-P. 1993, \apj, 407, 33

\bibitem[Miralda-Escude 1995]{ME95}
Miralda-Escud\'e, J. 1995, \apj, 438, 514

\bibitem[Moore, et al. (1998)]{1998ApJ...499L...5M} Moore, B., Governato, 
F., Quinn, T., Stadel, J. \& Lake, G. 1998, \apjl, 499, L5 

\bibitem[Moore, Lake, Quinn and Stadel (1999)]{1999MNRAS.304..465M} 
Moore, B., Lake, G., Quinn, T. \& Stadel, J. 1999a, \mnras, 304, 465 

\bibitem[Moore, et al. (1999)]{1999MNRAS.310.1147M} Moore, B., Quinn, T., 
Governato, F., Stadel, J. \& Lake, G. 1999b, \mnras, 310, 1147 

% \bibitem[Narayan \& Bartelmann 1998]{NB98} Narayan, R. \& Bartelmann, M. 1998,
%  in Proc.\ 1995 Jerusalem Winter School, Formation of Structure in the Universe,
%  ed.\ A. Dekel \& J. P. Ostriker (Cambridge Univ.~Press)

\bibitem[Navarro, Frenk, \& White 1996]{NFW96}
Navarro, J. F., Frenk, C. S., \& White, S. D. M. 1996, \apj, 462, 563

\bibitem[Navarro, Frenk and White (1997)]{1997ApJ...490..493N}
Navarro, J. F., Frenk, C. S. and White, S. D. M. 1997, \apj, 490, 493 

\bibitem[Oegerle and Hoessel (1991)]{1991ApJ...375...15O} 
Oegerle, W. R. \& Hoessel, J. G. 1991, \apj, 375, 15 

\bibitem[Oke \etal\ 1995]{Oke95}  Oke, J. B., Cohen, J. G., Carr, M.,
  Cromer, J., Dingizian, A., Harris, F. H., Labrecque, S., Lucinio, R.,
  Schaal, W., Epps, H., \& Miller, J. 1995, PASP, 107, 307

\bibitem[PHG]{PHG92} Postman, M. Huchra, J. P., \& Geller, M. J. 1992, \apj, 384, 404

\bibitem[PL]{pl} Postman, M. \& Lauer, T.\ R. 1995, ApJ, 440, 28 

\bibitem[Schlegel, Finkbeiner, \& Davis 1998]{SFD98}
Schlegel, D., Finkbeiner, D., \& Davis, M. 1998,  \apj, 500, 525

\bibitem[Shapley 1934]{Shap34} Shapley, H. 1934, \mnras, 94, 53

\bibitem[Smail \etal\ 1995]{Smail95}  Smail, I., Hogg, D. W., Blandford, R.,
 Cohen, J. G., Edge, A. C., \& Djorgovski, S. G. 1995, \mnras, 277, 1

\bibitem[Struble \& Rood 1987]{sr87} Struble, M. F. \& Rood, H. J. 1987, \apjs, 63, 555
\bibitem[Struble \& Rood 1999]{sr99} Struble, M. F. \& Rood, H. J. 1999, \apjs, 125, 35

\bibitem[Tarenghi \etal\ 1980]{Tar80}
Tarenghi, M., Chincarini, G., Rood, H. J., \& Thompson, L. A. 1980, \apj, 235, 724

\bibitem[Tarenghi, et al. (1979)]{1979ApJ...234..793T} Tarenghi, M., Tifft, 
W. G., Chincarini, G., Rood, H. J. \& Thompson, L. A. 1979, \apj, 234, 793 

\bibitem[Tyson \etal\ 1998]{TKD98}
Tyson, J. A., Kochanski, G. P., \& Dell'Antonio, I. P. 1998 \apj, 498 L107

\bibitem[Waxman \& Miralda-Escude 1995]{WME95}
Waxman, E. \& Miralda-Escude, J. 1995, ApJ, 451, 451

\bibitem[Wegner, et al. (1999)]{1999MNRAS.305..259W} 
Wegner, G., Colless, M., Saglia, R. P., McMahan, R. K., Davies, R. L., 
Burstein, D. \& Baggley, G. 1999, \mnras, 305, 259 

\bibitem[White, Jones, \& Forman 1997]{WJF97}
White, D. A., Jones, C., \& Forman, W. 1997, \mnras, 292, 419

\bibitem[Williams, Navarro, \& Bartelmann 1999]{WNB99}
Williams, L. L. R., Navarro, J. F., \& Bartelmann, M. 1999, \apj, 527, 535

\bibitem[Zabludoff \etal\ 1993]{zab93}
Zabludoff, A. I., Geller, M. J., Huchra, J. P., \& Ramella, M. 1993, \aj, 106, 1301

\end{thebibliography}
\end{document}